 \documentclass[final,5p,times,twocolumn]{elsarticle}

\usepackage{amssymb}

\usepackage{natbib}
\biboptions{square, comma, sort&compress}

\journal{Physics Letters B}

\begin{document}

\begin{frontmatter}



\title{Effect of final state interactions on particle production in $d$+Au collisions at RHIC}


\author[nju,lbl]{Xiaoping Zhang\corref{mail}}
\ead{\ xpzhangnju@gmail.com} \cortext[mail]{\ Corresponding
author}
\author[sinap]{Jinhui Chen}
\author[nju]{Zhongzhou Ren}
\author[lbl]{N. Xu}
\author[bnl]{Zhangbu Xu}
\author[nju,guizhou]{Qiang Zheng}
\author[qinghua]{Xianglei Zhu}
\address[nju]{Department of Physics, Nanjing University,
Nanjing 210008, China}
\address[lbl]{Nuclear Science Division, Lawrence Berkeley National
Laboratory, Berkeley, CA 94720, USA}
\address[sinap]{Shanghai Institute of Applied Physics, CAS, Shanghai, 201800, China}
\address[bnl]{Brookhaven National Laboratory, Upton, New York, 11973,
USA}
\address[guizhou]{School of Mathematics and Computer Science, Guizhou Normal
University, Guiyang, 550001, China}
\address[qinghua]{Department of Engineering Physics, Tsinghua University,
Beijing 100084, China }
\begin{abstract}
We show that particle species dependence of enhanced hadron
production at intermediate transverse momentum ($p_{T}$) for
$d$+Au collisions at RHIC can be understood in terms of the
hadronic rescatterings in the final state. A multiphase transport
model (AMPT) with two different hadronization mechanisms: string
fragmentation or parton coalescence, is used in our study. When
the hadrons are formed from string fragmentation, the subsequent
hadronic rescatterings will result in particle mass dependence of
nuclear modification factor $R_{CP}$, which is consistent with the
present experimental data. On the other hand, in the framework of
parton coalescence, the mass dependence disappears and the
strangeness plays an important role. Both mechanisms failed to
reproduce the $p_{T}$ dependence of $R_{CP}$ of pion, indicating
that initial-state effects might be also important in such
collisions.
\end{abstract}


\end{frontmatter}


\section{Introduction}
\label{} The Cronin effect, which refers to the enhanced hadron
production at intermediate transverse momentum ($p_{T}$) with
increasing target nucleus size in proton-nucleus (pA) collisions,
was first observed by Cronin {\it et al.} \cite{Cronin1,Cronin2}
in 1975 at center of mass energy $\sqrt{s_{NN}}$ = 27.4 GeV.
Recent experimental data in $d$+Au collisions from RHIC have shown
that similar effect exists at higher collision energy ($\sqrt{s}$
= 200 GeV) \cite{rdau1,rdau2}. An adequate understanding of the
Cronin effect becomes especially important for making reliable
theoretical interpretations of the observations in heavy ion
collisions at RHIC, in which the quark-gluon plasma (QGP) is
thought to be created \cite{qgp}.
Traditional explanations of the Cronin effect all involve multiple
scattering of incoming partons that lead to an enhancement at
intermediate $p_{T}$ \cite{theo1,theo2,theo3,theo4,theo5,theo6}.
The models can reproduce the observed centrality dependence for
pions very well. However, none of these initial-state models would
predict a species-dependent Cronin effect, as initial state parton
scattering precedes fragmentation into the different hadronic
species \cite{rdau2}. This indicates possible final-state
interactions (FSI) would possibly contribute to the Cronin effect.

Recently, Hwa and Yang \cite{rudy1} demonstrated the recombination
of soft and shower partons in the final state could explain the
mass dependent Cronin effect. This model do predict a larger
enhancement for protons than for pions at $1<p_{T}<4$ GeV/c.
However, the inclusion of recombination from deconfined partons
requires a high energy density in initial state, which may not be
justified in $d$+Au collisions.
An alternate way for hadronization is string fragmentation
\cite{hijing,ampt1,ampt2}. In this scenario, hadrons are formed
from the decays of excited strings, which result from the
recombination of energetic minijet partons and soft strings that
produced from initial soft nucleon-nucleon interactions. After the
hadronization, the followed rescatterings between formed hadrons
or between formed hadrons and nucleon spectators should also be
taken into account.

In this paper we study quantitatively how the two different
hadronization mechanisms (string fragmentation and parton
coalescence) and the followed final state interactions would
contribute to the nuclear modification factors in $d$+Au
collisions at $\sqrt{s}$ = 200 GeV. A multiphase transport model
(AMPT) \cite{ampt1,ampt2} with two versions: default
(hadronization from Lund string fragmentation, mainly hadronic
rescatterings in the final state, version 1.11) and string melting
(hadronization from quark coalescence, version 2.11) are used to
study the later-stage effect. The final state hadronic and/or
partonic interactions are included in the calculations. Quark
transverse momentum kick due to multiple scatterings are treated
in the same way as in Ref. \cite{hijing}. To have a clean test of
final state interactions, we assume no quark intrinsic $p_{T}$
broadening due to initial multiple parton scattering \cite{theo3},
and see how the final state interactions would contribute to the
Cronin effect observed. We show that recent data on particle
species dependence of central-to-peripheral nuclear modification
factor $R_{CP}$ at midrapidity for $d$+Au collisions at RHIC can
be understood in terms of the hadronization from string
fragmentation and the followed hadronic rescatterings in the final
state.

\section{The AMPT model}
The AMPT model \cite{ampt1,ampt2} is a hybrid model that consists
of four main components: the initial conditions, the partonic
interactions, conversion from the partonic to the hadronic matter,
and the hadronic interactions. The initial conditions, which
include the spatial and momentum distributions of hard minijet
partons and soft string excitations, are obtained from the HIJING
model (version 1.383 for this study). One uses a Woods-Saxon
radial shape for the colliding gold nuclei and introduces a
parameterized nuclear shadowing function that depends on the
impact parameter of the collision. The ratio of quark structure
function is parameterized as the following
impact-parameter-dependent but $Q^{2}$ (and flavor)-independent
form \cite{ampt1}
\begin{eqnarray}
R_{A}(x,r)\equiv\frac{f_{a}^{A}(x,Q^{2},r)}{Af_{a}^{N}(x,Q^{2})} \
\ \ \ \  \ \ \ \ \ \ \ \ \ \ \ \ \ \ \ \ \ \ \ \ \ \ \ \ \ \ \
\nonumber\\
=1+1.19\ln^{1/6}A(x^{3}-1.2x^{2}+0.21x)\ \ \nonumber\\
\
-[\alpha_{A}(r)-\frac{1.08(A^{1/3}-1)\sqrt{x}}{\ln(A+1)}]e^{-x^{2}/0.01},
\end{eqnarray}
where $x$ is the light-cone momentum fraction of parton a, and
$f_{a}$ is the parton distribution function. The impact-parameter
dependence of the nuclear showing effect is controlled by
\begin{eqnarray}
\alpha_{A}(r)=0.133(A^{1/3}-1)\sqrt{1-r^{2}/R^{2}_{0}},
\end{eqnarray}
with $r$ denoting the transverse distance of an interacting
nucleon from the center of the nucleus with radius
$R_{0}=1.2A^{1/3}$.
 The structure of
deuteron is described by the Hulthen wave function.
Scatterings among partons are modeled by the Zhang's parton
cascade (ZPC) \cite{zpc}, which at present includes only two-body
elastic scatterings with cross sections obtained from the pQCD
with screening masses. In the default AMPT model, after partons
stop interacting, they recombine with their parent strings, which
are produced from initial soft nucleon-nucleon interactions. The
resulting strings are converted to hadrons using the Lund string
fragmentation model. In case of string melting, the produced
hadrons from string fragmentation, are converted instead to their
valence quarks and antiquarks. The followed partonic interactions
are modeled by ZPC. After the partons freeze out, they are
recombined into hadrons through a quark coalescence process. The
dynamics of the subsequent hadronic matter is described by a
hadronic cascade, which is based on a relativistic transport model
(ART). Final hadronic observables including contributions from the
strong decays of resonances are determined when the hadronic
matter freezes out.
\begin{figure}[htbp]
\centering \vspace{-0.2cm}
\includegraphics[width=9cm]{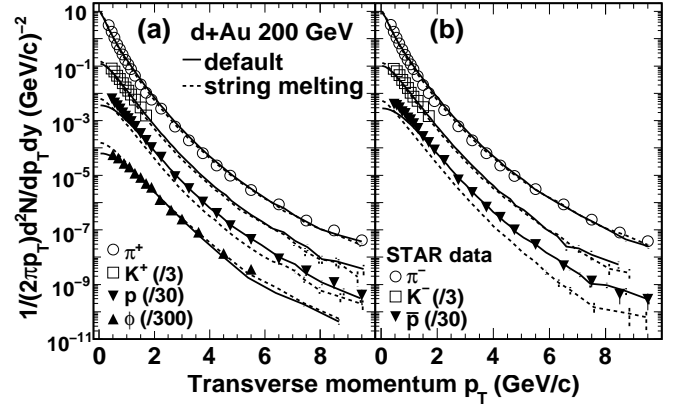}
\caption{\label{fig1.fig}Transverse momentum spectra of
mid-rapidity ($|y|<0.5$) pions, kaons, protons and $\phi$ mesons
in $``$minimum bias" $d$+Au collisions from default AMPT (solid
lines) and string melting AMPT (dashed lines) versus data from
STAR Collaboration (statistical error only) \cite{rdau1,phi}.}
\end{figure}

\section{Results}
We learn that Lin and Ko have done a nice study \cite{ampt2} on
global properties of deuteron-gold collisions with default AMPT
model, which shows good agreement with later experimental data
\cite{etaasym,fwlambda}. Their study on nuclear effects is up to
$p_{T}=2$ GeV/c. Here we focus on the intermediate to higher
$p_{T}$ range where the Cronin effect exists. We study the
deuteron-gold collisions at $\sqrt{s_{NN}}=200$ GeV.
The string fragmentation parameters are chosen to be the same as
in Ref. \cite{ampt2}. The partonic scattering cross section is
chosen to be 3 mb. The events are separated into different
centrality bins using the number of participant nucleons suffering
inelastic collisions. Fig. 1 shows the mid-rapidity ($|y|<$0.5)
transverse momentum spectra of pions, kaons, protons and $\phi$
mesons for $``$minimum bias" $d$+Au collisions from default AMPT
(solid line) and string melting (dashed line). It is seen that
both default and string melting AMPT can reproduce the $\pi^{\pm}$
and $K^{\pm}$ spectra well. For proton and antiproton production,
the default version works well for $p_{T}>1$ GeV/c, but
underestimates the low $p_{T}$ proton and antiproton yields. The
string melting version underestimates the proton and antiproton
production in the whole $p_{T}$ range.
For $\phi$ meson spectrum, the default version works well in the
whole $p_{T}$ range, while the string melting one overestimates
the low $p_{T}$ $\phi$ meson yields in the $``$minimum bias"
$d$+Au collisions.
\begin{figure}[htbp]
\centering
\includegraphics[width=9cm]{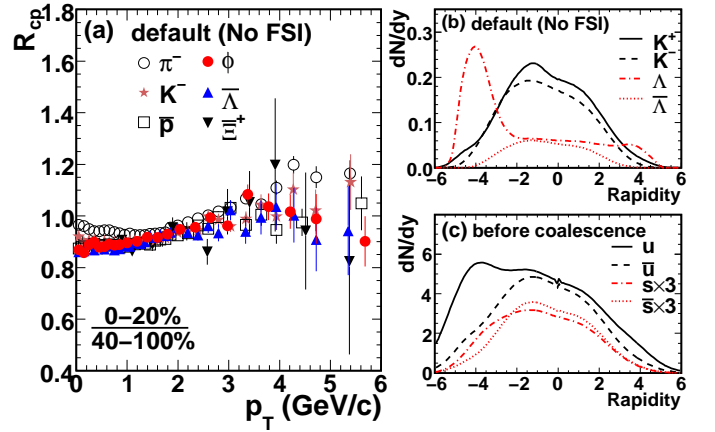}
\caption{\label{fig2.fig} (Color online) (a) $R_{CP}$ for
$\pi^{-}$, $K^{-}$, $\bar{p}$, $\phi$, $\overline{\Lambda}$ and
$\overline{\Xi}^{+}$ in $d$+Au $\sqrt{s_{NN}}$ = 200 GeV
collisions from default AMPT without final state interactions
included. (b) Kaon and $\Lambda$ rapidity distribution in
$``$minimum bias" $d$+Au collisions from default AMPT without
final state interactions included. (c) quark (u, $\bar{u}$, s, and
$\bar{s}$) rapidity distribution before coalescence in $``$minimum
bias" $d$+Au collisions from string melting AMPT.}
\end{figure}

To study the final state effect on the nuclear modification factor
$R_{CP}$, we first calculate the $R_{CP}$ (0-20\%/40-100\%
centrality) of different hadrons without including the final state
hadronic interactions and resonance decays in the default AMPT.
The $R_{CP}$, which compares particle yield from central
collisions to that of peripheral collisions, is defined as the
ratio of particle yields in central collisions over those in
peripheral ones scaled by number of inelastic binary collisions
$N_{\textrm{bin}}$, that is,
\begin{eqnarray}
R_{CP}=\frac{[dN/(N_{bin}p_{T}dp_{T})]_{\textrm{central}}}{[dN/(N_{bin}p_{T}dp_{T})]_{\textrm{peripheral}}}.
\end{eqnarray}
Here we use the same $N_{bin}$ value as STAR Collaboration at the
corresponding collision centrality \cite{rdau1}. One can see in
Fig. 2(a) that there are only slight differences for the $R_{CP}$
of different particle species. This is because hadrons are
produced from string fragmentation in the Lund model, and the
fragmentation patterns for different particle species in central
collisions and in peripheral collisions are set to be the same. We
note that the $R_{CP}$ of proton and $\Lambda$ would be larger due
to the associate production $N$+$N\rightarrow N$+$\Lambda$+$K^{+}$
from initial-state multiple scatterings in the gold beam direction
($y<0$) at large rapidity, as shown in Fig. 2(b) the enhanced
production of $\Lambda$ and $K^{+}$ in $``$minimum bias" $d$+Au
collisions. Some of the particles are scattered into mid-rapidity
region. The strange quark enhancement in the large rapidity region
(gold beam direction) will cause the corresponding increase of
$\bar{s}$ quark at other rapidity regions due to net strangeness
conservation. This is shown in Fig. 2(c): the quark rapidity
distribution before coalescence in $``$minimum bias" $d$+Au
collisions from the string melting AMPT.
\begin{figure}[htbp]
\centering
\includegraphics[width=8.7cm]{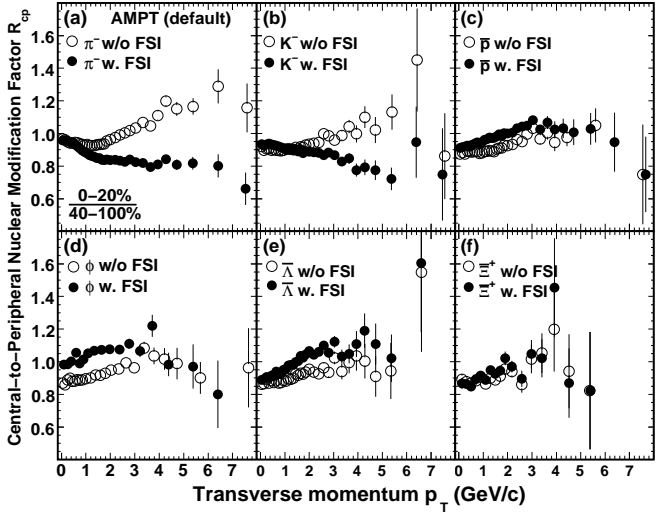}
\caption{\label{fig3.fig} Mid-rapidity $R_{CP}$ ($|y|<0.5$) in
$d$+Au $\sqrt{s_{NN}}$ = 200 GeV collisions from default AMPT with
(red circles) and without (black circles) final state
interactions.}
\end{figure}

After including the final-state hadronic interactions and strong
decays of resonances, the $R_{CP}$ of different particle species
will change differently, as they have different masses and
scattering cross sections. We show in Fig. 3 the comparisons of
$R_{CP}$ with and without the final state rescatterings and
resonance decays. For $\pi^{-}$ and $K^{-}$, the $R_{CP}$
decreases after including the final state interactions from
intermediate to high $p_{T}$. And this suppression increases with
$p_{T}$. Since most of the produced particles are pions at
mid-rapidity, the scatterings are mainly the particle-pion
interactions for $p\gg m_{0}$, where $m_{0}$ is the mass of
corresponding particles. For $\pi-\pi$ elastic collisions, the
resonance peak centers at the position of $\pi-\pi$ center-of-mass
energy $\sqrt{s_{\pi\pi}}$ close to $\rho$ meson rest mass. Since
most of the outgoing particles which probably scatter with each
other are in the similar directions, the open angle between two
scattering particles is small.
In our studied $p_{T}$ and rapidity range, for one particle at low
$p_{T}$, and another particle with higher energy, the calculated
$\sqrt{s_{\pi\pi}}$ is closer to the resonance peak. As a result,
this hadronic effect on $R_{CP}$ is enhanced with $p_{T}$ for
pion. Similar argument is also valid for K$-\pi$ scatterings. For
heavier particles like antiproton, $\phi$ meson,
$\overline{\Lambda}$, the $R_{CP}$ increases at low $p_{T}$
according to their corresponding production channels or due to
diffusions into mid-rapidity region, but changes slightly for
$p_{T}>3$ GeV/c. For even heavier particle, like
$\overline{\Xi}^{+}$, there is no obvious change of $R_{CP}$ due
to the final state interactions. Here particle mass plays
important roles in the hadronic rescatterings. It will determine
the space-time configuration of formed hadrons \cite{spacetime}.




\begin{figure}[htbp]
\centering
\includegraphics[width=9.2cm]{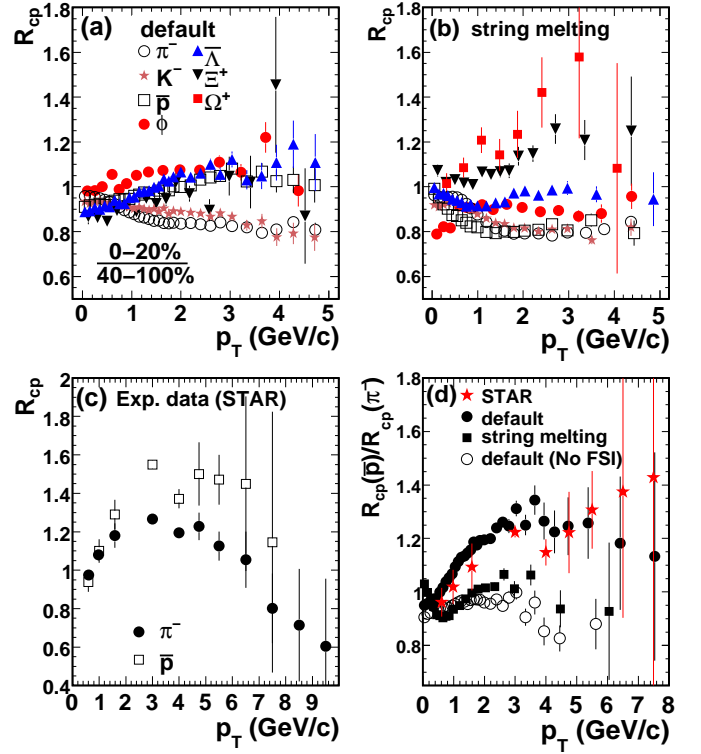}
\caption{\label{fig3.fig} (Color online) Mid-rapidity ($|y|<0.5$)
$R_{CP}$ in $d$+Au $\sqrt{s_{NN}}$ = 200 GeV collisions: (a)
default AMPT with final state interactions; (b) string melting
AMPT with final state interactions; (c) experimental data of
$R_{CP}$ from STAR Collaboration (statistical error only)
\cite{rdau1}. (d) The ratios of $R_{CP}(\bar{p})/R_{CP}(\pi^{-})$
from STAR Collaboration, from default AMPT, from string melting
AMPT, and from default AMPT without final state interactions.}
\end{figure}

According to above analysis, the final state hadronic
rescatterings will lead to particle mass dependence of $R_{CP}$.
In Fig. 4, we compare the data with model calculations. At
intermediate $p_{T}$, the $R_{CP}$ of heavier particles like
antiproton, $\phi$ meson, $\overline{\Lambda}$,
$\overline{\Xi}^{+}$ will be larger than those of $\pi^{-}$ and
$K^{-}$. The result is qualitatively consistent with experimental
data \cite{rdau1}, as shown in Fig. 4(c). At intermediate $p_{T}$,
the $R_{CP}$ of antiproton is systematically larger than that of
$\pi^{-}$. In Fig. 4(d), the ratio
$R_{CP}$($\bar{p}$)/$R_{CP}$($\pi^{-}$) from the default AMPT
model also agrees very well with experimental data. We note that
the present calculation without including the possible quark
intrinsic $p_{T}$ broadening at initial state can not reproduce
the $p_{T}$ dependence of $R_{CP}$. This difference between data
and model results may indicate that the initial state effect is
important.
The year 2008 data of RHIC with higher statistics will provide
more precise measurements and test our predictions for other
hadron species like $\phi$ meson, $\overline{\Lambda}$,
$\overline{\Xi}^{+}$, etc.

For comparisons, the $R_{CP}$ from string melting AMPT with quark
coalescence is also studied. We have shown in Fig. 2(c) that the
excess of $\bar{s}$ quark over $s$ quark at mid-rapidity is partly
due to associate production from initial multiple interactions.
Combining this effect with the coalescence of partons, there are
enhancements of corresponding hadrons at intermediate $p_{T}$. The
$R_{CP}$ values for different particle species that contain
different number of $\bar{s}$-quarks are shown in Fig. 4(b). Note,
multi-strange hadrons are particularly interesting as they suffer
much less hadronic interactions \cite{nxuplot} compared with
non-strange hadrons. Therefore they are more sensitive to early
stage dynamics. At intermediate $p_{T}$, there is an enhancement
of $R_{CP}$ according to the number of $\bar{s}$ quarks, that is,
$R_{CP}(\overline{\Lambda})<R_{CP}(\overline{\Xi}^{+})<R_{CP}(\overline{\Omega}^{+})$.
Note that the values of $R_{CP}$ for strange particles ($\Lambda$,
$\Xi$, and $\Omega$) are close to each other (not shown here) at
the same transverse momentum region. If one assumes the validity
of the coalescence approach, this observation shows that the
measured $R_{CP}$ can, to some extend, reflect the density of
quarks shortly before the freeze-out. However, from the
coalescence calculation the $R_{CP}$ of antiproton is close to
that of $\pi^{-}$ at intermediate $p_{T}$,
which is not consistent with experimental data, as shown in Fig.
4(d).


\section{Summary}
In summary, we studied the mechanism of hadron formation and
subsequent interactions in $d$+Au collisions at $\sqrt{s}$ = 200
GeV. In a multiphase transport model with Lund string
fragmentation for hadronization and the subsequent hadronic
rescatterings included, we find particle mass dependence of
central-to-peripheral nuclear modification factor $R_{CP}$.
Recent data on particle species dependence of $R_{CP}$ at
mid-rapidity for $d$+Au collisions at RHIC can be understood in
terms of this final-state hadronic rescatterings. This shows the
importance of final state hadronic interactions in $d$+Au
collisions, since none of the initial-state models would predict a
species-dependent $R_{CP}$ at present.
However, the calculations can not reproduce the $p_{T}$ dependence
of $R_{CP}$ with only final state interactions, this indicates the
initial state effects might be also important.
Issues associated with the initial condition such as gluon
saturation \cite{sat1,sat2,sat3}, parton intrinsic $p_{T}$
broadening \cite{theo1,theo2,theo3,theo4,theo5,theo6} and so on
were not addressed in this paper. The future more complete
analysis should include these effects. In comparison, if the
hadron is formed from quark coalescence, it is difficult to
explain antiproton transverse momentum spectra and the particle
species dependence of $R_{CP}$. On the other hand, the strangeness
effect plays an important role. More precision data in the future
will test the findings in this paper.
\\
\\
\textbf{Acknowledgements}

\vspace{1.5ex}
We are grateful to X.
Dong, C. Jena, H. Masui, B. Mohanty, Z. Lin, L. Ruan, and Z. B.
Tang for valuable discussions. This work is supported by the U.S.
Department of Energy under Contract No. DE-AC03-76SF00098, the
National Natural Science Foundation of China (Grant Nos. 10535010,
10775123, 10865004, 10905029, and 10905085), by the 973 National
Major State Basic Research and Development of China (Grant No.
G2000077400), by the CAS Knowledge Innovation Project No.
KJCX2-SW-N02, and by the Research Fund of High Education under
contract No. 20010284036.


\vspace{3ex} \hspace{-2.4ex}\textbf{References} \vspace{1.5ex}



\end{document}